\documentstyle{article}
\begin{document}
\thispagestyle{empty}
\parskip=12pt
\raggedbottom

\def\mytoday#1{{ } \ifcase\month \or
 January\or February\or March\or April\or May\or June\or
 July\or August\or September\or October\or November\or December\fi
 \space \number\year}
\noindent
\hspace*{9cm} BUTP--92/28\\
\hspace*{9cm} COLO-HEP-289\\
\vspace*{1cm}
\begin{center}
{\LARGE The Equivalence of the SU$(N)$ Yang-Mills Theory with a Purely
Fermionic Model}
\footnote{Work supported in part by Schweizerischer Nationalfonds and
NSF Grant PHY-9023257}

\vspace{2cm}

Anna Hasenfratz
\footnote{Present address: Physics Dept., University of Colorado,
Boulder CO 80309-390}
\\
University of Arizona Tucson, Department of Physics
\\
Tucson, AZ 85721, USA

\vspace{1cm}

Peter Hasenfratz \\
Institute for Theoretical Physics \\
University of Bern \\
Sidlerstrasse 5, CH-3012 Bern, Switzerland

\vspace{1cm}

\mytoday \\ \vspace*{1cm}

\nopagebreak[4]

\begin{abstract}
We investigate the detailed conditions under which a purely fermionic model
with current-current interaction goes over to a renormalizable, asymptotically
free SU$(N)$ gauge theory.

\end{abstract}

\end{center}

\eject
         The idea of composite gauge particles has a long history. The
         obvious possibility to form composite massless vector
         particles through vector condensates [1] is plagued by
         observable Lorentz symmetry breaking [2]. On the other hand,
         tightly bound vector states, in the limit when their mass
         goes to zero, are expected to have gauge interactions at low
         energies. This problem has been investigated by many
         theorists [2-5] especially in the context of composite $W$
         and $Z$ bosons in electroweak interactions [6]. These
         investigations are phenomenological in the sense that the
         high energy content of these models is left
         non-renormalizable, or it is not investigated explicitly
         further.

         It has been demonstrated recently [7-9] that the purely
         fermionic generalized Nambu-Jona-Lasinio model is completely
         equivalent to a perturbatively renormalizable quantum field
         theory which is formulated in terms of elementary scalars
         and fermions. Although the analytic calculations in [7] refer
         to the large-$N_{\mbox{\tiny{colour}}}$ limit, the field
         theoretical basis of the arguments is independent of $N_c$.
         In this letter we discuss the problem of constructing the
         standard renormalizable SU$(N)$ Yang-Mills theory out of a
         purely fermionic model for arbitrary $N \geq 2$. The physical
         content of the model is the usual gauge invariant
         selfinteraction of vector gluons. For
         this reason -- rather than talking about bound states and
         compositeness -- one can consider the original fermionic
         model as a specific realization of the Yang Mills theory.
         Admittedly, the point of view taken in [7] and here is
         somewhat different from that of the original idea of
         composite Higgs [10] or vector bosons [6]. The equivalence of
         purely fermionic models and theories with elementary scalars
         and gauge bosons does not have, in itself, phenomenological
         consequences. We find it intriguing, nevertheless, that the
         full standard model might be formulated in terms of fermion
         fields only.

         It is easy to understand, how a fermionic theory with
         4-fermion interaction might lead to a local gauge theory.
         Consider the free Lagrangean of $N_c$ massive Dirac fermions
         \begin{equation}
         {\cal L} (x) = \bar{\psi}^a (x) (\gamma^\mu \partial_\mu + M)
         \psi^a (x) \hspace{0.5cm}, \hspace{0.5cm} a = 1, 2, \ldots,
         N_c.
         \end{equation}
         This model has a U$(N_c)$ global symmetry with the
         corresponding conserved currents. Let us demand that the
         SU$(N_c)$ Noether currents are pointwise zero,
         i.e.
         \begin{equation}
         j^A_\mu(x) = \bar{\psi} (x) \gamma^\mu T^A \psi (x) = 0, \;
         \forall x,\; \mu = 1, \ldots,d,\; A = 1, \ldots, N^2_c-1,
         \end{equation}
         where $T^A$ are the SU$(N_c)$ generators.
         The model defined by eqs. (1), (2) has a {\em local}
         SU$(N_c)$ symmetry. Indeed, the variation of $\cal L$ under a
         local SU$(N_c)$ transformation is proportional to the current
         $j^A_\mu (x)$ which is zero according to eq. (2). On the
         other hand, the constraint $\delta (j^A_\mu (x))$ in the path
         integral can be represented by a 4-fermion current-current
         interaction using $\delta (y) \sim \lim_{\Delta \rightarrow 0}
         e^{- \frac{1}{\Delta} y^2}$.

         In order to investigate further under which circumstances
         will be the corresponding gauge invariant model a cut-off
         independent Yang-Mills theory, we shall introduce a
         regularization and generalize the model slightly. We shall
         use dimensional regularization and introduce $N_f$ fermion
         fields: $\psi^a_i (x)$, where $a= 1, \ldots, N_c$ and $ i= 1,
         \ldots , N_f$ refer to colour and flavour, respectively. As
         we discussed before, the Lagrangean
         \begin{equation}
         {\cal L} = \bar{\psi} (\gamma^\mu \partial_\mu + M) \psi +
         \frac{\kappa^\epsilon}{2m^2}\; (\bar{\psi} \gamma^\mu T^A \psi)
         \; (\bar{\psi} \gamma^\mu T^A \psi)
         \end{equation}
         in the limit $m \rightarrow 0$, $\epsilon$ fixed, defines a
         model with local SU$(N_c)$ symmetry in $n= 4 - \epsilon$
         dimension. Here $\kappa$ is a finite arbitrary mass scale.
         Introducing the auxiliary fields $W^A_\mu (x)$ one obtains in
         the $m \rightarrow 0$ limit
         \begin{equation}
         \int D \psi D \bar{\psi} DW_\mu \;\exp \left \{ - \int_{x}
         \left [ \bar{\psi} \left ( \gamma^\mu \partial_\mu + M \right
         ) \psi- i \kappa^{\epsilon/2} \;W^A_\mu \bar{\psi} \gamma^\mu
         T^A \psi \right ] \right. .
         \end{equation}
         The gauge symmetry in eq. (4) is obvious: eq. (4) is the
         standard, gauge invariant fermion-gluon interaction, but the
         gluon part $\sim F^A_{\mu \nu} F^A_{\mu \nu}$ is missing.
         Integrating over the fermions we get
         \begin{equation}
         \int DW_\mu \;\exp \left \{ - S(W_\mu) \right \},
         \end{equation}
         where the gauge field action $S(W_\mu)$ is a sum over
         one-loop graphs with $l$ gluon legs, $l= 2,3,\ldots$:
         \bigskip
         \begin{equation}
	 \vspace{.7cm}
         \end{equation}
         \bigskip
         The first term in eq. (6), giving the quadratic part of the
         action has the form
         \begin{equation}
         \frac{1}{e^2_0} \;\int_{p} \frac{1}{2}\; W^A_\mu(p) W^A_\nu (p)
         (p^2 \delta_{\mu\nu} - p_\mu p_\nu)\; 6\; \int^{1}_{0} dx\; x (1-x)
         \left [1 + x(1-x) \frac{p^2}{M^2} \right]^{-\epsilon/2},
         \end{equation}
         with
         \begin{equation}
         e_0  =  \bar{e}_0 \kappa^{-\epsilon/2},
         \hspace{1.0cm}
         \frac{1}{\bar{e}^2_0}  =  N_f M^{-\epsilon} \;\frac{2}{3}\;
         \frac{1}{(4 \pi)^{n/2}}\; \Gamma \left (2- \frac{n}{2}\right ).
         \end{equation}
         For momenta $p^2/M^2 \ll 1$, eq. (7) corresponds to
         $1/4e^2_0 \int_x (\partial_\mu W^A_\nu (x) -
         \partial_\nu W^A_\mu (x))^2$. The graphs with 3 and 4 legs in
         eq. (6) make this expression gauge invariant, the square
         of the standard field strength tensor will enter. The bare
         gauge coupling $\bar{e}^2_0$ is proportional to $\epsilon$,
         goes to zero as the regularization is removed, as it should
         in an asymptotically free theory. The one-loop graphs with
         more than 4 legs in eq. (6) are convergent and are suppressed
         by $1/M$ powers for momenta much below $M$.

         In order to proceed further, we have to make a digression and
         discuss two questions in the {\em standard} dimensionally
         regularized Yang-Mills theory.
         \newline
         A.{\it The relation between the
            bare coupling and $\Lambda_{\mbox{\tiny{MS}}}$}
         \newline
         Let us consider the action
         \begin{equation}
         S^{\mbox{\tiny{YM}}} = \frac{1}{4} \int F^A_{\mu \nu}
         F^A_{\mu \nu} d^nx,
         \end{equation}
         where
         \begin{equation}
         F^A_{\mu \nu} = \partial_\mu A^A_\nu-\partial_\nu A^A_\mu
                        - \bar{g}_0 C^{ABC}
         A^B_\mu A^C_\nu,
         \end{equation}
         and $\bar{g}_0$ is the bare coupling, $\dim \bar{g}_0 =
         (\mbox{mass})^{-\epsilon/2}$. The precise way the bare
         coupling $\bar{g}_0$ goes to zero as $\epsilon \rightarrow 0$
         will determine the typical scale ($\Lambda$-parameter) of the
         resulting continuum Yang-Mills theory. Let us tune $\bar{g}_0 =
         \bar{g}_0 (\epsilon)$ towards zero as
         \begin{equation}
         \frac{1}{\bar{g}^2_0}\; = \;\left ( \frac{2 \beta_0}{\epsilon} +
         \frac{\beta_1}{\beta_0} \ln \frac{1}{\epsilon} \right )
         \Lambda^{- \epsilon},
         \end{equation}
         where $\beta_0 $ and $\beta_1 $ are the first two universal
         coefficients of the $\beta$-function and $\Lambda$ is some
         arbitrary mass. Then the typical low energy scale of the
         resulting Yang-Mills theory in the $\epsilon \rightarrow 0$
         limit will be $\sim \Lambda$. For example, the scale
         parameter in the MS-scheme is related to $\Lambda$ as
         \begin{equation}
         \Lambda_{\mbox{\tiny{MS}}} = \;\exp \left
         (-\frac{\beta_1}{2\beta_0^2} (1-\ln 2)\right)  \Lambda.
         \end{equation}
         Eq. (12) can be derived as follows. On the two-loop level,
         eqs. (9), (10) lead to the following relation between the
         renormalized coupling in the MS-scheme $g_{\mbox{\tiny{MS}}}
         (\mu)$ and the bare coupling $\bar{g}_0$:
         \begin{equation}
         g^2_{\mbox{\tiny{MS}}} (\mu) = \bar{g}^2_0 \mu^{-\epsilon} +
         \left ( \bar{g}^2_0 \mu^{-\epsilon} \right )^2
         \frac{2\beta_0}{\epsilon} + \left ( \bar{g}^2_0
         \mu^{-\epsilon} \right )^3 \left ( \frac{\beta_1}{\epsilon} +
         \frac{4\beta^2_0}{\epsilon^2} \right ).
         \end{equation}
         Eq. (13) contains pole terms only (MS-scheme) and leads to
         the correct $\beta$-function
         \begin{equation}
         \mu \frac{d}{d\mu} g^2_{\mbox{\tiny{MS}}} (\mu) = - \epsilon
         g^2_{\mbox{\tiny{MS}}} (\mu) - 2 \beta_0
         g^4_{\mbox{\tiny{MS}}} (\mu) - 2 \beta_1
         g^6_{\mbox{\tiny{MS}}} (\mu) + \ldots .
         \end{equation}
         Integrate the renormalization group equation eq. (14) between
         $\mu$ and $\mu^\prime$, where $\mu^\prime$ is so large that
         $g^2_{\mbox{\tiny{MS}}} (\mu^\prime)\ll \epsilon$, and so --
         according to eq. (13 ) -- $g^2_{\mbox{\tiny{MS}}} (\mu^\prime)
         = \bar{g}^2_0 \mu^{\prime -\epsilon}$. This way one
         finds the relation between $\bar{g}^2_0$ and
         $g^2_{\mbox{\tiny{MS}}} (\mu)$ which reads
         \begin{equation}
         \frac{1}{\bar{g}^2_0 \mu^{-\epsilon}} =
         \frac{2\beta_0}{\epsilon} + \frac{1}{g^2_{\mbox{\tiny{MS}}}
         (\mu)} - \frac{\beta_1}{\beta_0} \left [ \ln \left (
         \frac{\epsilon}{2 \beta_0 g^2_{\mbox{\tiny{MS}}} (\mu)}
         \right ) + 1 \right ].
         \end{equation}
         Eqs. (11), (15) and the two-loop definition
         \begin{equation}
         \Lambda_{\mbox{\tiny{MS}}} = \mu \left ( \beta_0
         g^2_{\mbox{\tiny{MS}}} (\mu \right)^{-\beta_1/2 \beta^2_0}
         \exp \left \{ - \frac{1}{2 \beta_0 g^2_{\mbox{\tiny{MS}}}
         (\mu)} \right \}
         \end{equation}
         lead to the relation eq. (12).
         \newline
         B. {\it Large $M$ limit at fixed regularization}
         \newline
         Let us add to the action eq. (9) some small gauge
         invariant perturbation which modifies the quadratic part and
         the vertices. The perturbation can be local, or nonlocal. We
         assume that the perturbation depends on some mass scale $M$ in
         such a way that for any set of fixed momenta flowing into the
         bare vertices or propagator the perturbation goes to zero as $M
         \rightarrow \infty$. Consider this modified model and take
         the limit $M \rightarrow \infty$ at {\em fixed} value of the
         regularization parameter. One expects that the effect of the
         perturbation disappears in this limit, the Green's functions
         remain unchanged. Consider any graph before and after the
         perturbation is introduced and take the difference. For any
         fixed set of internal and external momenta the integrand of
         the corresponding momentum integrals goes to zero as $M
         \rightarrow \infty$. If the regularization is such that the
         region of momentum integration is constrained (lattice, for
         example) then the integral itself goes to zero also, and the
         intuitive expectation is satisfied. The situation
         is less obvious if dimensional regularization is used and,
         actually, we are not able to present a formal proof. We shall
         illustrate on the example of the 1-loop $\beta$-function that the
         $M \rightarrow \infty$ limit at fixed regularization parameter
         $\epsilon$ (the precise conditions are given in eqs. (18),(19))
         leads to the standard Yang-Mills result, as expected. We shall
         assume that this is true in every order of perturbation theory.

         We return now to the 4-fermion formulation and investigate
         the structure of the vertices in eq. (5) in more detail.
         Introduce the rescaled gauge field $A^A_\mu (x) = 1/e_0
         W^A_\mu (x)$ and consider a vertex with $l$ external
         $A$-lines
         \bigskip
         \begin{equation}
	 \vspace{1.0cm}
         \end{equation}
         \bigskip
         Here
         $V^{A_1 \ldots A_l}_{\mu_1 \ldots \mu_l}
         (p_i, M) = \bar{e}^2_0 N_f \cdot
         f^{A_1 \ldots A_l}_{\mu_1 \ldots \mu_l}
         (p_i,M)$, where $f$ is a sum of 1-loop
         integrals
         with $l$ massive fermion propagators. Since $\bar{e}^2_0 \sim
         1/N_f$ (eq. (8)), $V$ is independent of $N_f$. Counting
         Lorentz indices and dimensions we get that for $ p \ll M$ the
         leading behaviour of $V$ is $\sim M^{4-l}$ for $l =$ even
         and $ \sim pM^{3-l}$ if $l = $ odd. The
         action in eq. (5) has therefore the following properties: (i)
         The $l$-point vertex is $\bar{e}_0^{l-2}
         V^{A_1 \ldots A_l}_{\mu_1 \ldots \mu_l}
         (p_i, M)$, where $V$ is independent of $N_f$. (ii)
         For $p \ll M$, only the 3- and 4-point vertices survive and
         their form agrees with that of the standard Yang-Mills
         vertices. (iii) The quadratic part of the action has the
         standard Yang-Mills form for $p \ll M$.

         Consider now the limit at fixed regularization parameter
         $\epsilon = 4-n$:
         \begin{equation}
         N_f \rightarrow \infty\;, \; M \rightarrow \infty, N_f \cdot
         M^{-\epsilon} = c( \epsilon)\;, \;\mbox{fixed}
         \footnote{i.e. the bare charge is fixed},
         \end{equation}
         where $c(\epsilon)$ is chosen so that the bare charge in eq.
         (8) has the form
         \begin{equation}
         1/\bar{e}^2_0 = \left ( \frac{2 \beta_0}{\epsilon} +
         \frac{\beta_1}{\beta_0} \;\ln\; \frac{1}{\epsilon} \right )
         \Lambda^{- \epsilon}
         \end{equation}
         with some arbitrary, fixed scale $\Lambda$. As we discussed in
         point B, (i)-(iii) and eq. (18) imply that in this limit the
         standard Yang-Mills model is obtained in $n=4-\epsilon$
         dimension with a bare coupling $\bar{e}^2_0$. Eq. (19)
         assures then, according to our discussion in point A, that
         in the $\epsilon \rightarrow 0$ limit a continuum, cut-off
         independent Yang-Mills theory is obtained whose low energy
         scale is $\sim \Lambda$. Eqs. (18), (19) give [11]
         \begin{equation}
         M = \Lambda\;\epsilon^{\beta_1/ 2\beta^2_0}\;
         \exp\left (\frac{1}{\epsilon}
         \ln \frac{2N_f}{11 N_c} +\frac{1}{2}(\Gamma^\prime (1)+ \ln 4
         \pi)\right )\; [1 + O(\epsilon)],
         \end{equation}
         showing that for any fixed $\epsilon$, $M/ \Lambda$ goes
         to infinity as $N_f \rightarrow \infty$.

         As an illustration, let us discuss the 1-loop
         $\beta$-function of the model in eq. (5). Since the action $S$
         is gauge invariant, the gauge fixing with the Fadeev-Popov
         procedure goes through as usual. In Landau gauge, for
         example, we get for the gluon propagator
         \begin{equation}
         D^{AB}_{\mu \nu} (q) = \delta_{AB} \left ( \delta_{\mu \nu} -
         \frac{q_\mu q_\nu}{q^2} \right ) \frac{1}{q^2 \cdot f(q^2)},
         \end{equation}
         where
         \begin{equation}
         f(q^2) = 6 \int^{1}_{0} dx \;x (1-x) \left [ 1 + x(1-x)
         \frac{q^2}{M^2} \right ]^{-\epsilon/2}.
         \end{equation}
         The ghost propagator and ghost vertex are standard. The
         1-loop graphs to the 2- and 3-point functions are also
         standard (although with the propagator eq. (21) and 3- and
         4-point vertices which become non-local at the scale of $M$),
         except a single graph in the 3-point function where two legs
         of the 5-point vertex is closed. We define the wave function
         renormalization and the renormalized coupling
         $e_{\mbox{\tiny{MOM}}} (\mu)$
         as usual


         \begin{equation}
         - \mu^2 \delta_{A_1A_2} \delta_{\mu_1 \mu_2} = Z\;
         \Gamma \begin{array}{c} (2) A_1 A_2 \\ \mu_1 \mu_2
         \end{array}(p) \mid_{p^2=\mu^2},
         \end{equation}
         $$
         ie_{\mbox{\tiny{MOM}}} (\mu) \mu^{\epsilon/2}\; C^{A_1A_2A_3}\;
         \left [\delta_{\mu_1\mu_3} (p_1 -p_3)_{\mu_2}
         + \delta_{\mu_1 \mu_2} (p_2 - p_1)_{\mu_3} + \delta_{\mu_2
         \mu_3} (p_3 -p_2)_{\mu_1}\right ]
         $$
         \begin{equation}
         = Z^{3/2} \Gamma \begin{array}{c} (3) \prime A_1A_2A_3 \\
         \mu_1 \mu_2 \mu_3 \end{array}
         (p_1,p_2,p_3)\mid_{\mbox{\tiny{S.P.}}}
         \end{equation}
         where $\Gamma^{(3)\prime}$ denotes the part of $\Gamma^{(3)}$
         which has the same Lorentz structure as the standard
         Yang-Mills vertex, while the symmetric point (S.P.) is
         defined by $p^2_i = \mu^2$, $p_ip_j = - \frac{1}{2} \mu^2$,
         $i \neq j$. The renormalized coupling is dimensionless. On
         dimensional ground we have
         \begin{equation}
         e_{\mbox{\tiny{MOM}}} (\mu)\mu^{\epsilon/2}
                  = \bar{e}_0 + \bar{e}^3_0\;
         M^{-\epsilon} \;f \left (\frac{\mu^2}{M^2}, \epsilon \right ).
         \end{equation}
         On the left hand side of eq. (25) $\mu$ enters only through
         the definition (see the left hand side of eq. (24)). On the
         right hand side of eq. (25) $\mu$ enters through the
         $\mu$-dependence of the graphs. Since $\mu/M \rightarrow 0$
         ($\epsilon$ fixed), it is clear that the right hand side of
         eq. (25) can have $\mu$-dependence only through infrared
         divergencies of the graphs for $\mu \rightarrow 0$. To find
         these infrared divergencies we can consider small loop
         momentum. Then, however, we have the standard vertices and
         propagators (up to $p^2/M^2 \rightarrow 0$ corrections).
         Consequently, the $\mu$-dependence of eq. (25) is identical
         to that of the standard Yang-Mills theory:
         \begin{equation}
         e_{\mbox{\tiny{MOM}}} (\mu) \mu^{\epsilon/2} = \bar{e}_0 +
               \bar{e}^3_0
         \left [ \mu^{-\epsilon}\left ( \frac{\beta_0}{\epsilon} + a
         \right ) + \mu\mbox{-independent} \right ],
         \end{equation}
         where $a$ is the same finite constant one obtains in the
         standard Yang-Mills theory. On dimensional ground the
         $\mu$-independent part in eq. (26) should be proportional to
         $M^{-\epsilon}$
         \begin{equation}
         e_{\mbox{\tiny{MOM}}} (\mu) \mu^{\epsilon/2}\; =\; \bar{e}_0 +
         \bar{e}_0 \mu^{-\epsilon} \left [ \left ( \frac{\beta_0}{\epsilon} + a
         \right ) + \left ( \frac{\mu}{M} \right )^\epsilon \cdot
         b(\epsilon) \right ].
         \end{equation}
         In the limit $M \rightarrow \infty$, $\epsilon$ fixed (eq.
         (18)) the last term goes to zero in agreement with the
         general expectation (see the discussion in point B above).
         The relation between $e_{\mbox{\tiny{MOM}}} (\mu)$ and $\bar{e}_0$
         will have then the standard form and so will
         the 1-loop $\beta$-function
         $\mu \frac{d}{d\mu} \;e_{\mbox{\tiny{MOM}}} (\mu)$.

         Let us turn back to the form in eq. (3) and discuss an interesting
         possibility. After a Fierz transformation [12] one can
         introduce {\em colour singlet} auxiliary
         fields and after integrating over the fermions, $N_c$
         will enter as an overall factor only. We were
         not able to make further progress along this line, however.
\newcommand{\PL}[3]{{Phys. Lett.} {\bf #1} {(19#2)} #3}
\newcommand{\PR}[3]{{Phys. Rev.} {\bf #1} {(19#2)}  #3}
\newcommand{\NP}[3]{{Nucl. Phys.} {\bf #1} {(19#2)} #3}
\newcommand{\PRL}[3]{{Phys. Rev. Lett.} {\bf #1} {(19#2)} #3}
\newcommand{\PREPC}[3]{{Phys. Rep.} {\bf #1} {(19#2)}  #3}
\newcommand{\ZPHYS}[3]{{Z. Phys.} {\bf #1} {(19#2)} #3}
\newcommand{\ANN}[3]{{Ann. Phys. (N.Y.)} {\bf #1} {(19#2)} #3}
\newcommand{\HELV}[3]{{Helv. Phys. Acta} {\bf #1} {(19#2)} #3}
\newcommand{\NC}[3]{{Nuovo Cim.} {\bf #1} {(19#2)} #3}
\newcommand{\CMP}[3]{{Comm. Math. Phys.} {\bf #1} {(19#2)} #3}
\newcommand{\REVMP}[3]{{Rev. Mod. Phys.} {\bf #1} {(19#2)} #3}
\newcommand{\ADD}[3]{{\hspace{.1truecm}}{\bf #1} {(19#2)} #3}
\newcommand{\PA}[3] {{Physica} {\bf #1} {(19#2)} #3}
\newcommand{\JE}[3] {{JETP} {\bf #1} {(19#2)} #3}
\newcommand{\FS}[3] {{Nucl. Phys.} {\bf #1}{[FS#2]} {(19#2)} #3}


\end{document}